\documentclass[twocolumn,aps,showpacs,prb,tightenlines,amsmath,amssymb,superscriptaddress]{revtex4}

\usepackage{bm}
\usepackage{ulem}
\usepackage{graphicx}                          
\usepackage{bmpsize}             
\usepackage{amssymb}                                                             
\usepackage{amsmath}
\usepackage{colordvi}
\usepackage{calrsfs}

\begin{document} 

\title{Majorana bound states in a disordered quantum dot chain}

\author{P. Zhang}
\affiliation{CEMS, RIKEN, Saitama 351-0198, Japan}
\author{Franco Nori}
\affiliation{CEMS, RIKEN, Saitama 351-0198, Japan}
\affiliation{Department of Physics, The University of Michigan, Ann Arbor, Michigan 48109-1040, USA}
\date{\today}

\begin{abstract}
We study Majorana bound states in a disordered chain of semiconductor quantum
dots proximity-coupled to an s-wave superconductor. By calculating its topological quantum number, based on the
scattering-matrix method and a tight-binding model, we can identify the topological property of such an
inhomogeneous one-dimensional system. We study the robustness of Majorana bound
states against disorder in both the spin-independent terms (including the chemical
potential and the regular spin-conserving hopping) and the spin-dependent term,
i.e., the spin-flip hopping due to the Rashba spin-orbit coupling. We find that
the Majorana bound states are {\it not} completely immune to the spin-independent disorder, especially when the latter is
strong. Meanwhile, the Majorana bound states are relatively robust against spin-dependent disorder, as long as the spin-flip hopping is of uniform sign
(i.e., the varying spin-flip hopping term does not change its sign along the
chain). Nevertheless, when the disorder induces sign-flip in spin-flip
hopping, the topological-nontopological phase transition takes place in the low-chemical-potential region.  

\end{abstract}
\pacs{71.10.Pm, 74.78.Na}

\maketitle
\section{Introduction}
Majorana bound states (MBS)\cite{Kitaev131,Beenakker113} in solid-state systems are recently attracting increasing interest, both theoretically and
experimentally. Proposed by Kitaev more than ten years ago in a spinless toy model,\cite{Kitaev131} these zero-energy bound states are expected to exist
in several structures with spin, including nanowires with spin-orbit coupling (SOC) in proximity to a
superconductor (SC),\cite{Lutchyn077001,Oreg177002,asano104513} ferromagnetic atom chains on top of a SC,\cite{Choy195442} topological
insulator/SC hybrid structures,\cite{fu096407,Rakhmanov075141,cook201105,Akzyanov085409,Akzyanov-arxiv,jiang075438}
quantum dot (QD) chains with SC in adjacence,\cite{sau964,fulga045020,dai11188} as well as cold-atom systems.\cite{jiang220402} Experimentally, possible signatures of MBS have
been reported in nanowires,\cite{Mourik1003,das887,deng6414} atom
chains,\cite{Nadj-Perg602} and topological insulator/SC structures.\cite{Xu017001}

Majorana bound states attract considerable attention partly due to their future
potential applications in quantum
information.\cite{Beenakker113,nayak1083,Alicea412,Tewari010506} One attractive
possibility would be to construct Majorana qubits based on MBS.\cite{nayak1083} 
Majorana qubits, among various qubit candidates,\cite{wuReview,perge2010,zhang115417,li086805,buluta326,buluta74,you474} are supposed to be robust against local perturbations and hence
promising to store quantum information.\cite{nayak1083,mao174506,sau964}
Moreover, arbitrary qubit rotations are expected to be implemented, by means of
topologically-protected braiding operations\cite{Alicea412,Ivanov268} in
combination with other non-topological operations assisted by, e.g., nanomechanical
resonators.\cite{kovaklev106402,zhang115303} However, recent studies reveal that
the MBS are not completely robust against disorder in the Kitaev's spinless
model and in the systems with
spin.\cite{brouwer144526,DeGottardi146404,brouwer196804,hu165118,Pekerten00449,Klinovaja62}
Moreover, the Majorana qubits are not totally protected from
decoherence.\cite{rainis174533,schmidt085414,Goldstein205109,Budich121405}

Note that the studies investigating so far the effect of disorder on MBS focus solely on the spin-independent disorder,
without considering the spin-dependent one. In fact, the spin-dependent
disorder, e.g., the randomness in SOC, can be present
inevitably in many solid-state systems and play an important
role in the spin-related dynamics.\cite{glazov156602,glazov2157,zhang033015} Therefore, the
effect of spin-dependent disorder on the existence of MBS deserves to be investigated.

\begin{figure}[t] 
  {\includegraphics[width=8.5cm]{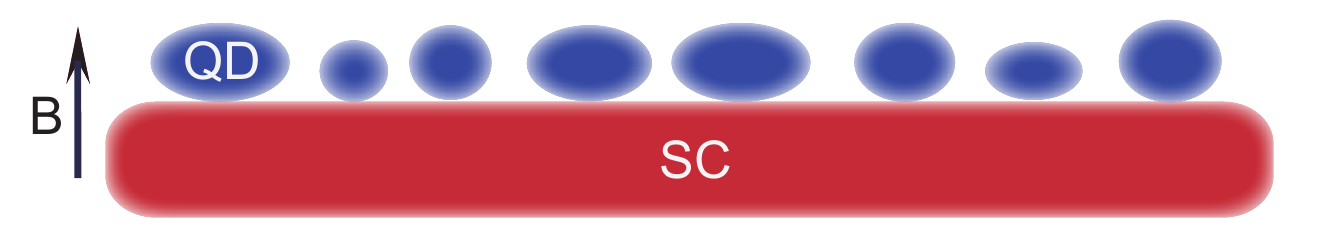}}
  \caption{(Color online) Schematic diagram of a disordered chain of
    semiconductor quantum dots (shown in blue) in proximity to an s-wave superconductor
    (in red), under a transverse magnetic field $B$. The on-site chemical
    potentials in the quantum dots, as well as the spin-conserving and -flip
    hopping terms between neighbouring quantum dots, can vary among the
    different sites. }
   \label{fig0}
\end{figure}

In this work, we systematically study the robustness of MBS against disorder, based on a
concrete structure, i.e., a QD chain in proximity to an s-wave superconductor.\cite{fulga045020}
Experimentally, such a QD chain system might have the advantage to be
adaptively tuned, as suggested in Ref.~\onlinecite{fulga045020}. However, in the absence of precise control,
this system is also very likely to be disordered due to, e.g., the inhomogeneity in QD sizes 
or QD confining potentials. Therefore, we consider a QD chain as an ideal
platform to study the influence of disorder. Concretely, we calculate the topological quantum number by means of the 
scattering-matrix method on a tight-binding model, to identify the topological property of a disordered
chain in a relatively large parameter region. Apart from the disorder in the
spin-independent terms (including the 
chemical potential and the regular spin-conserving hopping), we also consider
the disorder in the spin-dependent term, i.e., the spin-flip hopping due to the
Rashba SOC. We find that the MBS are {\sl not} completely immune to disorder in
the spin-independent terms, especially when the disorder is
strong. Meanwhile, the MBS are relatively robust against disorder in the spin-flip hopping, as long as the spin-flip hopping is 
of uniform sign. Nevertheless, when the disorder induces sign-flip in spin-flip
hopping, a topological-nontopological phase transition in the QD chain takes place in the low-chemical-potential region.

This paper is organized as follows. First, we describe the inhomogeneous QD
chain in a tight-binding model. Then we present the scattering-matrix method
used to calculate the topological quantum number. Afterwards, we 
numerically study the robustness of the MBS against disorder in the QD
chain. Finally, we summarize our results.

\section{Model and Hamiltonian}

A QD chain, as studied in Ref.~\onlinecite{fulga045020}, is schematically shown here in
Fig.~\ref{fig0}. An s-wave SC is in proximity to the QD chain
  and a transverse magnetic field $B$ is applied along the
  $z$-axis. We assume that the QDs can be approximately treated as one dimensional along the chain-direction ($x$-axis) due to the strong transverse
confinement. By further assuming that the orbital
level splitting in the QDs is much larger than both the Zeeman splitting and
Rashba SOC, we consider only the Kramers doublet closest to the
chemical potential energy in each QD.  The general form of the tight-binding Hamiltonian describing such a chain of single-level QDs is
written as\cite{fulga045020}
\begin{align}\nonumber
  H=&\frac{1}{2}\sum_{n\alpha\beta}[-\mu_n\delta_{\alpha\beta}+B(\sigma_{z})_{\alpha\beta}]f_{n\alpha}^\dagger
  f_{n\beta}+\Delta\sum_nf_{n\uparrow}^\dagger f_{n\downarrow}^\dagger \\ & +\sum_{n\alpha\beta}[t_n\delta_{\alpha\beta}+it^{\rm so}_n(\sigma_y)_{\alpha\beta}]f_{n\alpha}^\dagger
  f_{n+1\beta}+{\rm H.c.}.\label{ori-hami}
\end{align}
Here, $f^\dagger_{n\alpha}$ is the creation operator for a spin-$\alpha$ electron in the
$n$th QD. The Pauli matrices $\sigma_{x,y,z}$ act on the spin space. The chemical potential is labeled as $\mu_n$. The term proportional to $B$ is the Zeeman splitting while
$\Delta$ stands for the superconducting pairing due to the proximity effect. The nearest-neighbour hopping term has
two parts, i.e., the spin-conserving ($t_n$) and spin-flip ($t^{\rm so}_n$)
ones. The spin-flip hopping can be caused by the SOC which supplies an effective magnetic
field during hopping. Here we only consider the Rashba type SOC, with its
effective magnetic field along the $y$-axis. Due to the
inhomogeneity in the QD confining potentials and/or QD/SC sizes, as well as
other disorder sources such as charged impurities, both the spin-conserving terms, $\mu_n$ and $t_n$, and the
spin-flip term, $t_n^{\rm so}$, can be QD-site dependent. 

In the Bogoliubov-de Gennes basis $\Psi_n=(f_{n\uparrow},f_{n\downarrow},f_{n\downarrow}^\dagger,-f_{n\uparrow}^\dagger)$,
the Eq.~(\ref{ori-hami}) can be rewritten as\cite{Choy195442}
\begin{align}
  H=\frac{1}{2}\sum_n[\Psi_n^\dagger {\hat h}_n\Psi_n+(\Psi_n^\dagger {\hat t}_n\Psi_{n+1}+{\rm
      H.c.})],
  \label{hami}
\end{align}
where
\begin{align}
{\hat h}_n&=-\mu_n\sigma_0\tau_z+B\sigma_z\tau_0+\Delta\sigma_0\tau_x,\\
{\hat t}_n&=t_n\sigma_0\tau_z+i t^{\rm so}_n\sigma_y\tau_z, 
\end{align}
and the Pauli matrices $\tau_{x,y,z}$ act on the particle-hole space.

\section{Scattering-matrix method}
\label{method}
To identify the topological property of the QD chain, we study the scattering matrix $S$ relating the incoming and outgoing
wave amplitudes at the Fermi level\cite{akhmerov}

\begin{align}
  S=\left(
\begin{array}{cc}
R & T^\prime \\
T & R^\prime 
\end{array}
\right).
\label{scat}
\end{align}
The $4\times 4$ subblocks \{$R$, $R^\prime$\} and \{$T$, $T^\prime$\} are the reflection and transmission
matrices at the two ends of the QD chain, respectively. The $Z_2$ topological quantum number $Q$ is given by\cite{akhmerov}
\begin{align}\label{q}
  Q={\rm sgn\: Det}(R)={\rm sgn\: Det}(R^\prime).
\end{align}
Here, sgn denotes the sign of the determinant Det. The MBS arise\cite{akhmerov} at the ends of the QD chain only when $Q=-1$.

The scattering matrix can be obtained by the transfer-matrix scheme. Based on Hamiltonian (\ref{hami}), the zero-energy Schr\"odinger equation
gives\cite{Choy195442}
\begin{align}
\left(
\begin{array}{c}
  {\hat t}_n^\dagger\Phi_n \\
  \Phi_{n+1}
\end{array}
\right)={\tilde M}_n\left(
\begin{array}{c}
  {\hat t}_{n-1}^\dagger\Phi_{n-1}\\
  \Phi_n
\end{array}
\right),
\end{align}
where
\begin{align}
  {\tilde M}_n=\left(
\begin{array}{cc}
 0 & {\hat t}_n^\dagger\\
 -{\hat t}_n^{-1} & -{\hat t}_n^{-1}{\hat h}_n
\end{array}
\right).\label{mn}
\end{align}

Here $\Phi_n$ is a four-component vector of wave amplitudes on the $n$th site. The above recursive relation indicates that waves at the two ends ($n=1$
and $N$) of the nanowire are related by the transfer matrix 
\begin{align}
{\tilde M}={\tilde M}_N{\tilde M}_{N-1}...{\tilde M}_2{\tilde M}_1. 
\end{align}
In the basis with right-moving and left-moving waves
separated in the upper and lower four components, the transfer matrix transforms
as 
\begin{align}
  {M}_n=U^\dagger {\tilde M}_nU,\label{trans}
\end{align}
where
\begin{align}
  U=\frac{1}{\sqrt{2}}\left(
\begin{array}{cc}
 I & I\\
 iI & -iI
\end{array}
\right).
\end{align}
In this basis, 
the reflection matrices $R$ ($R^\prime$) and transmission matrices
$T$($T^\prime$) in the scattering matrix $S$ [refer to Eq.~(\ref{scat})] can be obtained via the relations
\begin{align}
\left(
\begin{array}{c}
  T \\
  0
\end{array}
\right)={M}\left(
\begin{array}{c}
  I\\
  R
\end{array}
\right),
\left(
\begin{array}{c}
  R^\prime \\
  I
\end{array}
\right)={M}\left(
\begin{array}{c}
  0\\
  T^\prime
\end{array}
\right),\label{R}
\end{align}
where
\begin{align}
{M}={M}_N{M}_{N-1}...{M}_2{M}_{1}.\label{tildem}
\end{align}

Finally, the calculation of the topological quantum number $Q$ is reduced to
that of the transfer matrix $M$. In Appendix~\ref{ap1}, we present the
numerical method for calculating $M$. 

\section{Results}
We now numerically study\cite{code} the topological property of the QD chain. For
comparison, we first look into an ideal homogeneous QD chain and reproduce the topological phase
reported in the literature, and then take into account disorder to
investigate the robustness of the MBS.

\subsection{Homogeneous QD chain}
\begin{figure}[th]
  {\includegraphics[width=7.2cm]{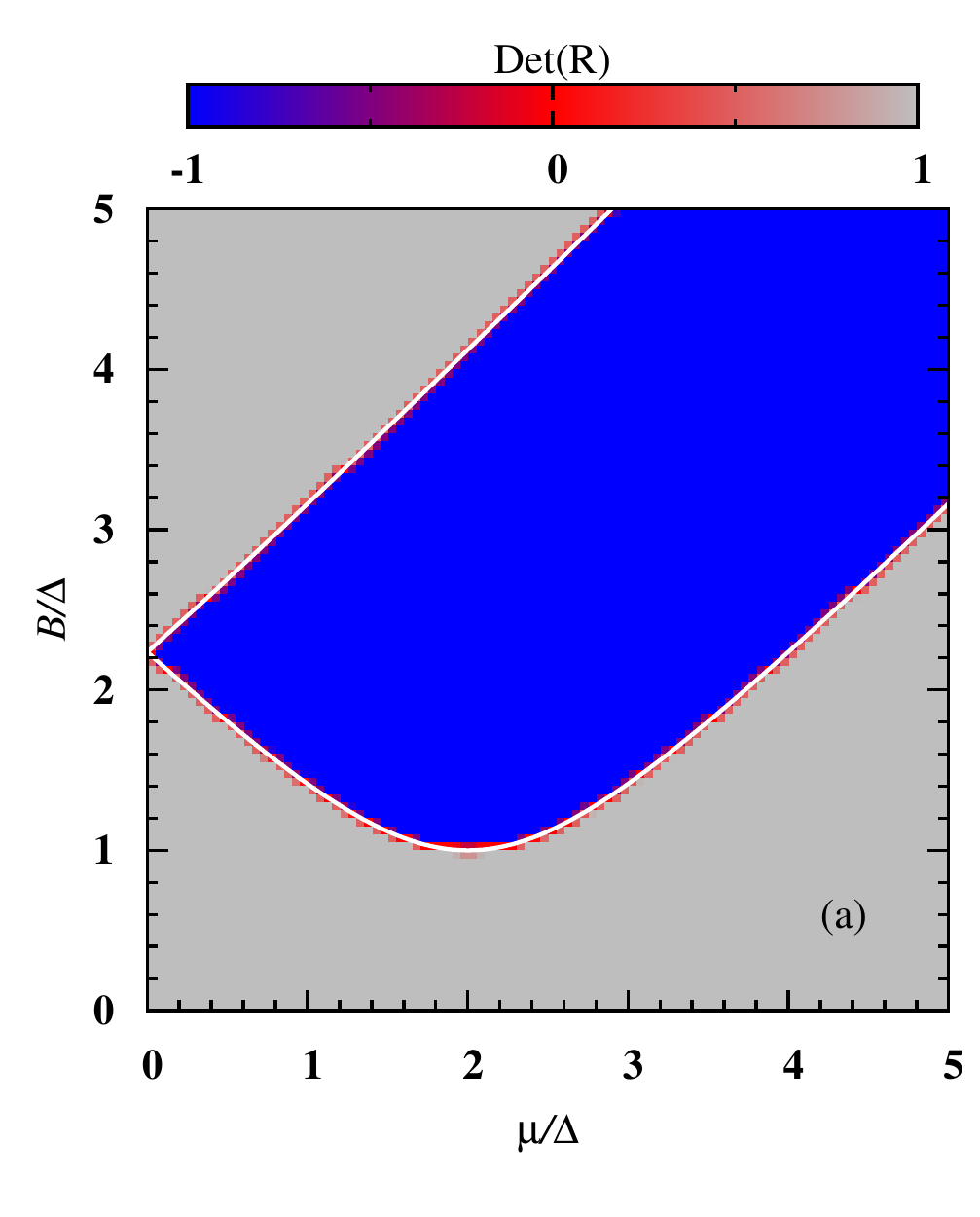}}\\{\includegraphics[width=7cm]{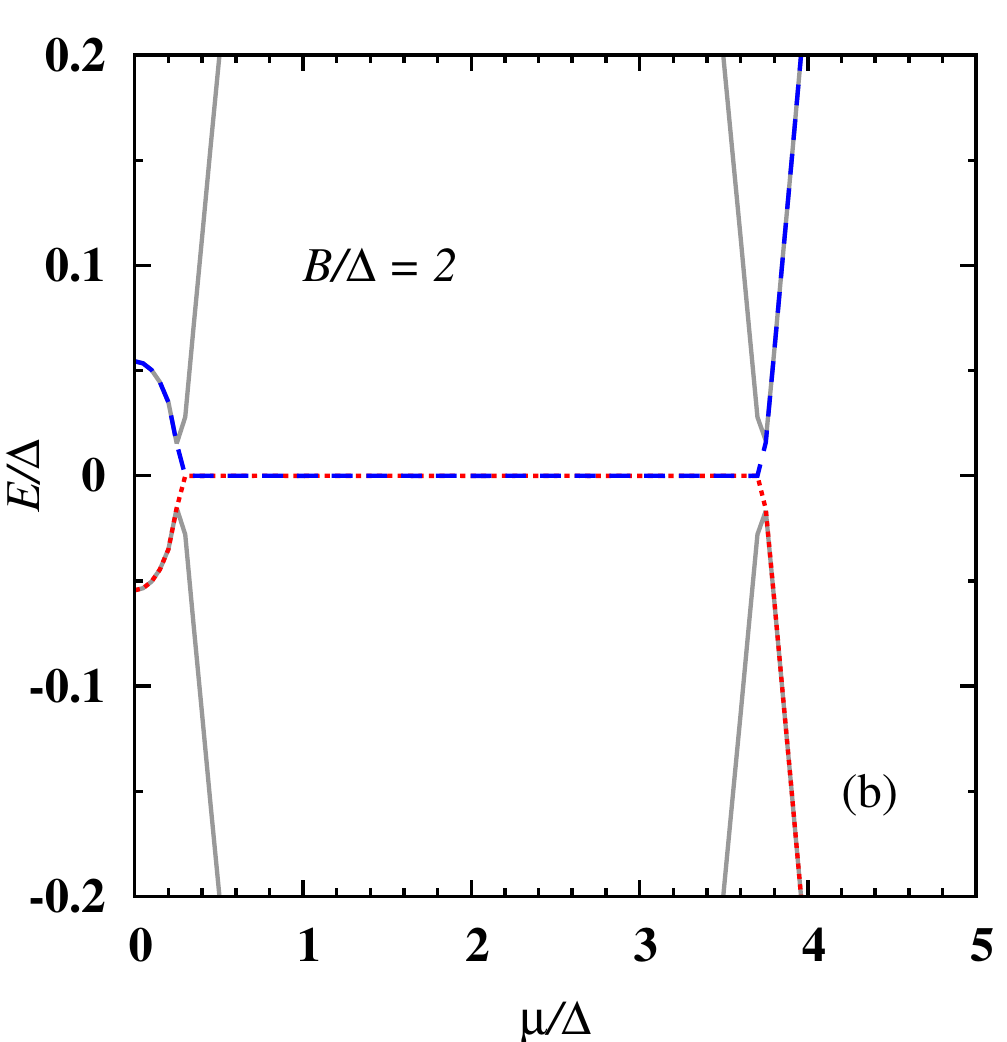}}
    \caption{(Color online) (a) The determinant ${\rm Det}(R)$ of the reflection matrix $R$ as a function of the chemical potential $\mu$ and the Zeeman splitting
      $B$, in a homogeneous QD chain with $t=\Delta$ and $t_{\rm so}=0.5\Delta$. The blue region with ${\rm Det}(R)=-1$ stands
      for the topological phase supporting MBS. (b) The energy spectrum (with
      only the lowest four eigenstates close to zero energy plotted)
      versus the chemical potential $\mu$, when the Zeeman splitting $B$ is
      fixed as $2\Delta$. Note that in this figure, as well as in Figs.~\ref{fig2}-\ref{fig3}, the chain has $N=500$ QDs, which is large enough for the numerical convergence.} 
    \label{fig1}
 \end{figure}

For a homogeneous QD chain, we denote $\mu_n=\mu$, $t_n=t$ and $t^{\rm so}_n=t_{\rm so}$. In Fig.~\ref{fig1}(a) we plot the phase
diagram, ${\rm Det}(R)$ [refer to Eqs.~(\ref{scat}) and (\ref{q})] versus $\mu$ and $B$, of a homogeneous
QD chain typically with $t=\Delta$ and $t_{\rm so}=0.5\Delta$. The blue region
in this figure, with ${\rm Det}(R)=-1$, stands for the topological phase supporting
MBS. It is found that this region is nicely enclosed by the white curve plotted in the figure, which
defines the topological region of a single-band homogeneous superconducting nanowire as\cite{fregoso180507,gibertini144525}
\begin{align} 
\sqrt{(2t-|\mu|)^2+\Delta^2}<|B|<\sqrt{(2t+|\mu|)^2+\Delta^2}.\label{region}
\end{align}
In Fig.~\ref{fig1}(b), we further show the energy spectrum (for clarity, we
present only the lowest four states close to the zero energy) of this QD chain
versus $\mu$ when $B$ is fixed. It is clear [from the red and blue curves in Fig.~2(b)] that when the QD chain
enters the topological region, the zero-energy states (localized at the two ends
of the QD chain) which are separated from the higher-energy bulk states arise. Note that when varying the spin-flip hopping
$t_{\rm so}$, the topological phase space in Fig.~\ref{fig1}(a) remains
invariant, consistent with the feature that $t_{\rm so}$ is absent from Eq.~(\ref{region}). 

\subsection{Inhomogeneous QD chain with disordered chemical potential and
  spin-conserving hopping}

\begin{figure}[th] 
  {\includegraphics[width=8cm]{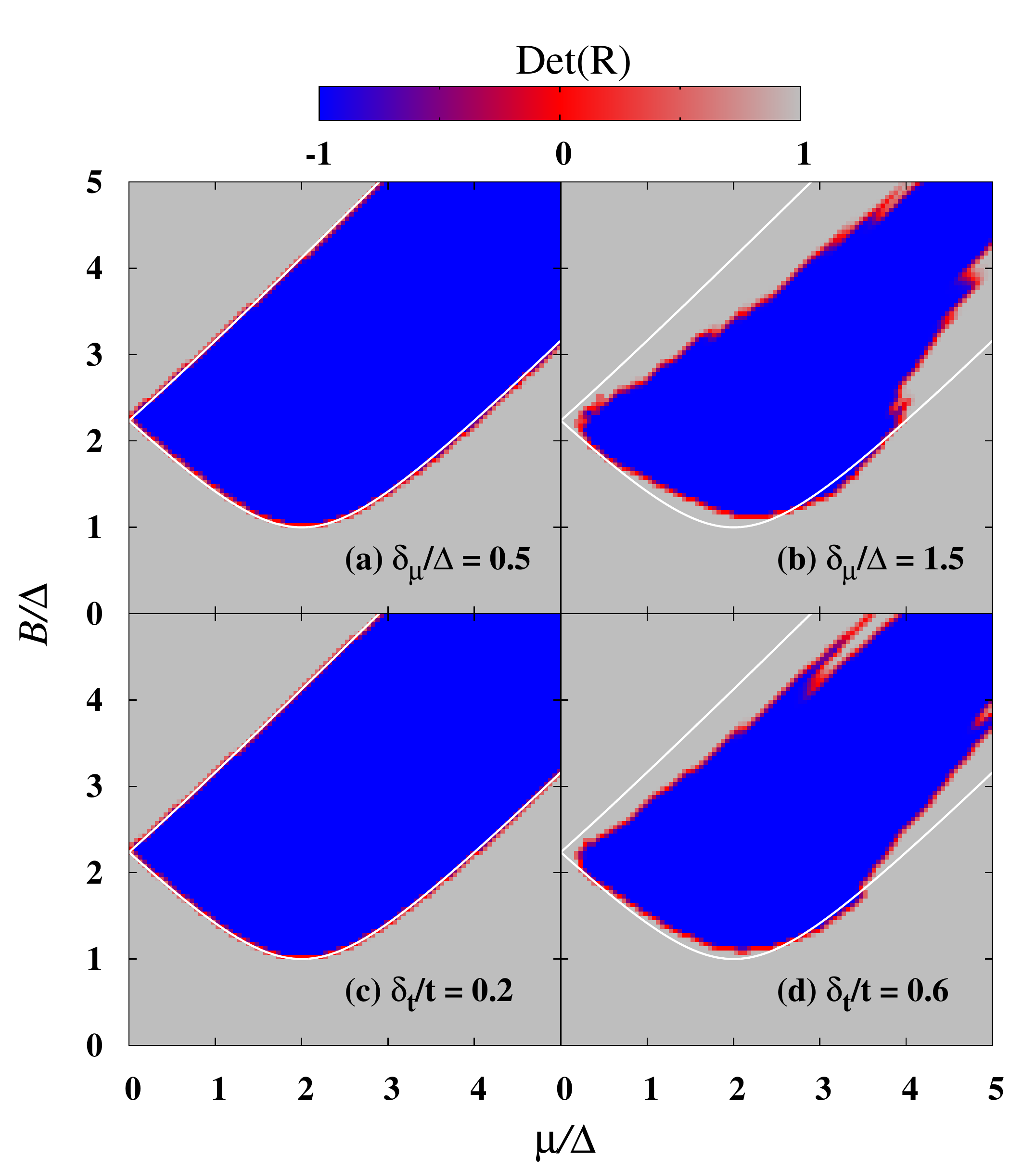}}
  {\includegraphics[width=5.5cm]{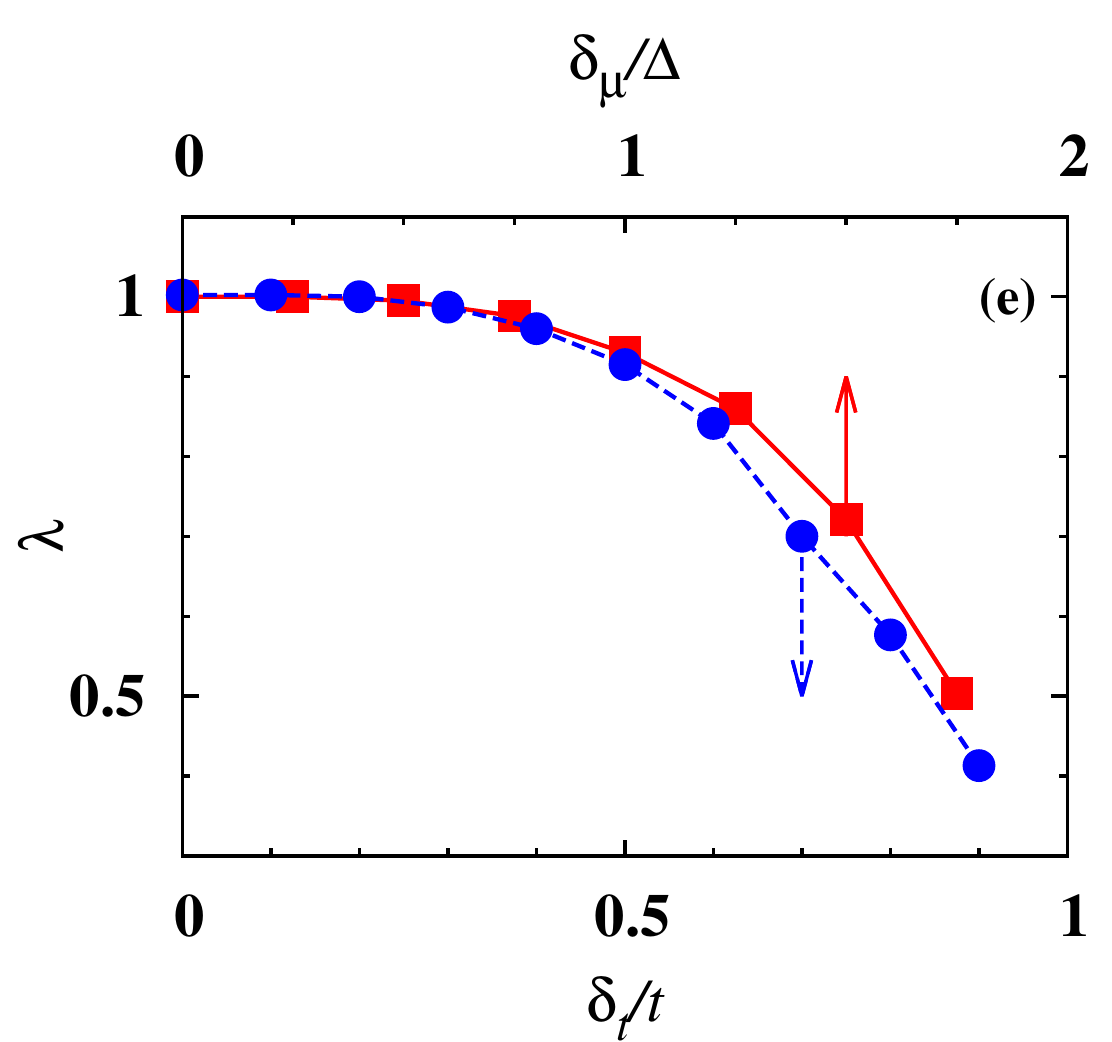}}
  \caption{(Color online) (a) and (b) [(c) and (d)] Phase diagrams of disordered QD chains,
    where the chemical potentials $\mu_n$ (spin-conserving hoppings $t_n$)
    fluctuate in an interval $(\mu-\delta_\mu,\mu+\delta_\mu)$
    [$(t-\delta_t,t+\delta_t)$] with a uniform distribution. Note that
    $\delta_\mu/\Delta$ is set as 0.5 and 1.5, respectively, in (a) and (b), and
    $\delta_t/t$ is set as 0.2 and 0.6, respectively, in (c) and (d). (e) The
    ratio of the area of the topological region for a disordered system [such as the blue regions in (a)-(d)] to the one for
    a clean system [the region defined by Eq.~(\ref{region}), or, enclosed by
      the white curves in (a)-(d)], labeled as $\lambda$, 
    versus the fluctuation magnitude $\delta_\mu$ of the chemical potential
    $\mu$ (red curve with squares), and the fluctuation magnitude $\delta_t$ of
    the spin-conserving hopping $t$ (blue curve with circles). The calculations for each curve in (e) are carried out by
    averaging over ten disordered samples.}  
   \label{fig2}
\end{figure}

From Eq.~(\ref{region}), one may infer that when the disorder is induced into
the chemical potential $\mu$ or the spin-conserving hopping $t_n$, the topological
phase space might change in the parameter space. Now we take into account such
disorder to investigate the robustness of MBS in the QD chain. We first
consider disorder in the chemical potential, which is modeled to perturb the $\mu_n$'s
independently within a uniform distribution in the interval
$(\mu-\delta_\mu,\mu+\delta_\mu)$, where $\mu$ is now the mean value of the
chemical potential and $\delta_\mu$ stands for the fluctuation magnitude. Our calculations indicate
that the topological phase is not completely immune to disorder. In
Figs.~\ref{fig2}(a) and (b), we present the phase diagrams of the inhomogeneous
QD chain calculated with $\delta_\mu/\Delta=0.5$ and $\delta_\mu/\Delta=1.5$, respectively. The comparison between these two figures
indicates the effect of stronger disorder on the formation of the topological phase. To
qualitatively present the effect of increasing disorder, we further study the ratio of the
area of the topological region with disorder [such as the blue regions in Figs.~\ref{fig2}(a)
and (b)] to that without disorder [the region defined by Eq.~(\ref{region})], labeled as $\lambda$, 
versus the fluctuation magnitude $\delta_\mu$. This is a qualitative study
because it is performed here in a finite parameter region, e.g., $0\leq \mu
\leq 5\Delta$ and $0\leq B\leq 5\Delta$. This result is shown by the solid curve with squares in Fig.~\ref{fig2}(e). This curve shows that when the fluctuation
magnitude of the chemical potential $\delta_\mu$ is larger than the superconducting
gap $\Delta$, the topological phase can be effectively destroyed.

We then consider disorder in the spin-conserving hopping, with the other terms
treated as uniform. We assume that the disorder causes the spin-conserving hopping to fluctuate in an interval $(t-\delta_t,t+\delta_t)$ with a uniform 
distribution ($\delta_t<t$). Our calculations indicate that disorder in the spin-conserving hopping can also be detrimental to the topological phase
(especially when the disorder is strong), as shown by the phase diagrams in
Figs.~\ref{fig2}(c) and (d). In Fig.~\ref{fig2}(e), by the blue curve with
circles, we also plot the ratio $\lambda$ of the area of the topological region
for a disordered system to the one for a clean system, versus the
fluctuation magnitude $\delta_t$. Also, the
stronger the disorder is, the smaller the topological phase area becomes.

\begin{figure}[t] 
  {\includegraphics[width=8cm]{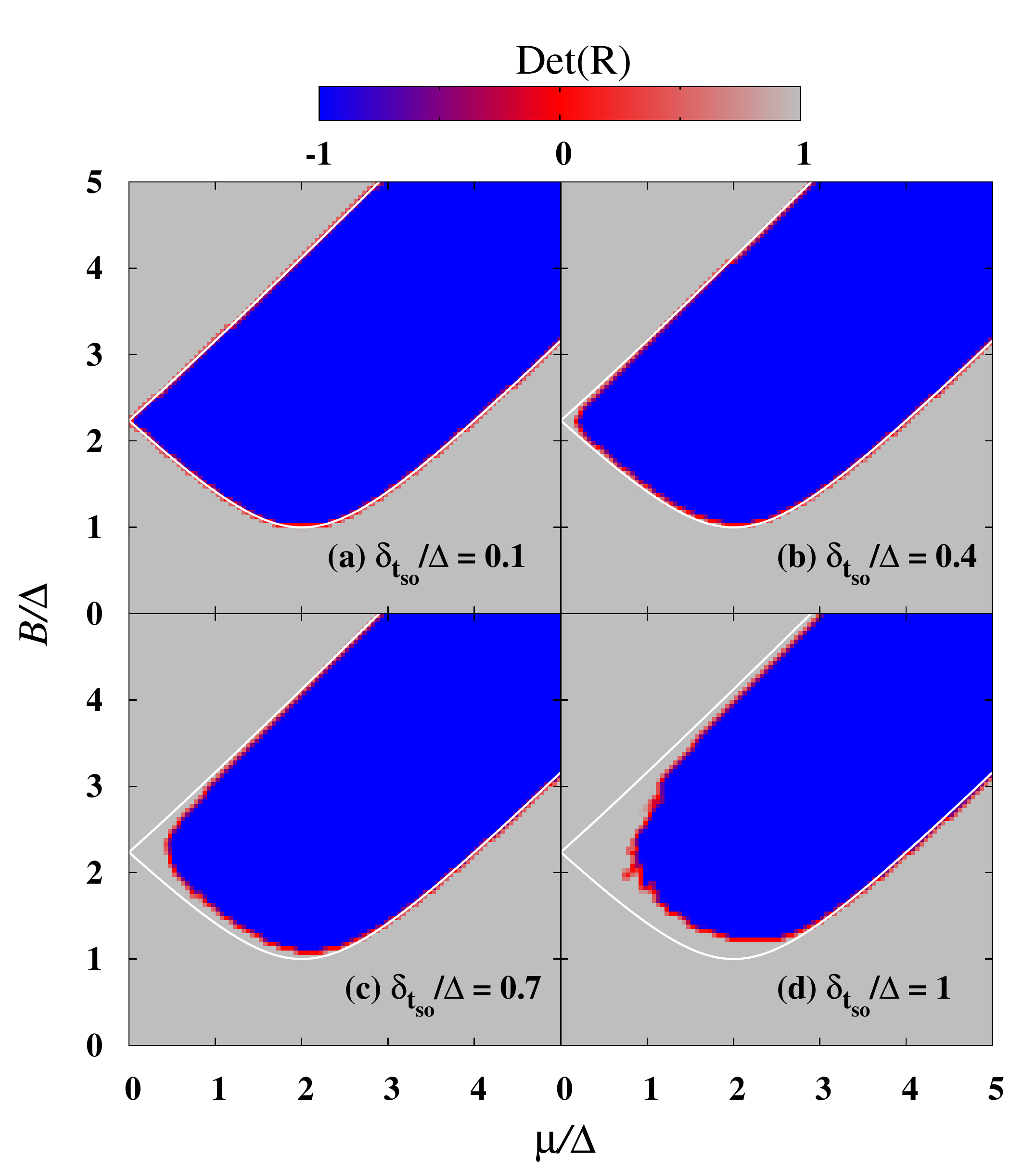}}
  \caption{(Color online) The phase diagrams of disordered QD chains
    where the spin-flip hoppings $t^{\rm so}_n$ fluctuate in an interval $(t_{\rm so}-\delta_{t_{\rm so}},t_{\rm so}+\delta_{t_{\rm
        so}})$ with a uniform distribution. The fluctuation magnitude
    $\delta_{t_{\rm so}}$ increases from (a) $0.1\Delta$ to (d) $\Delta$.}  
   \label{fig3}
\end{figure}

\begin{figure}[th] 
  {\includegraphics[width=9cm]{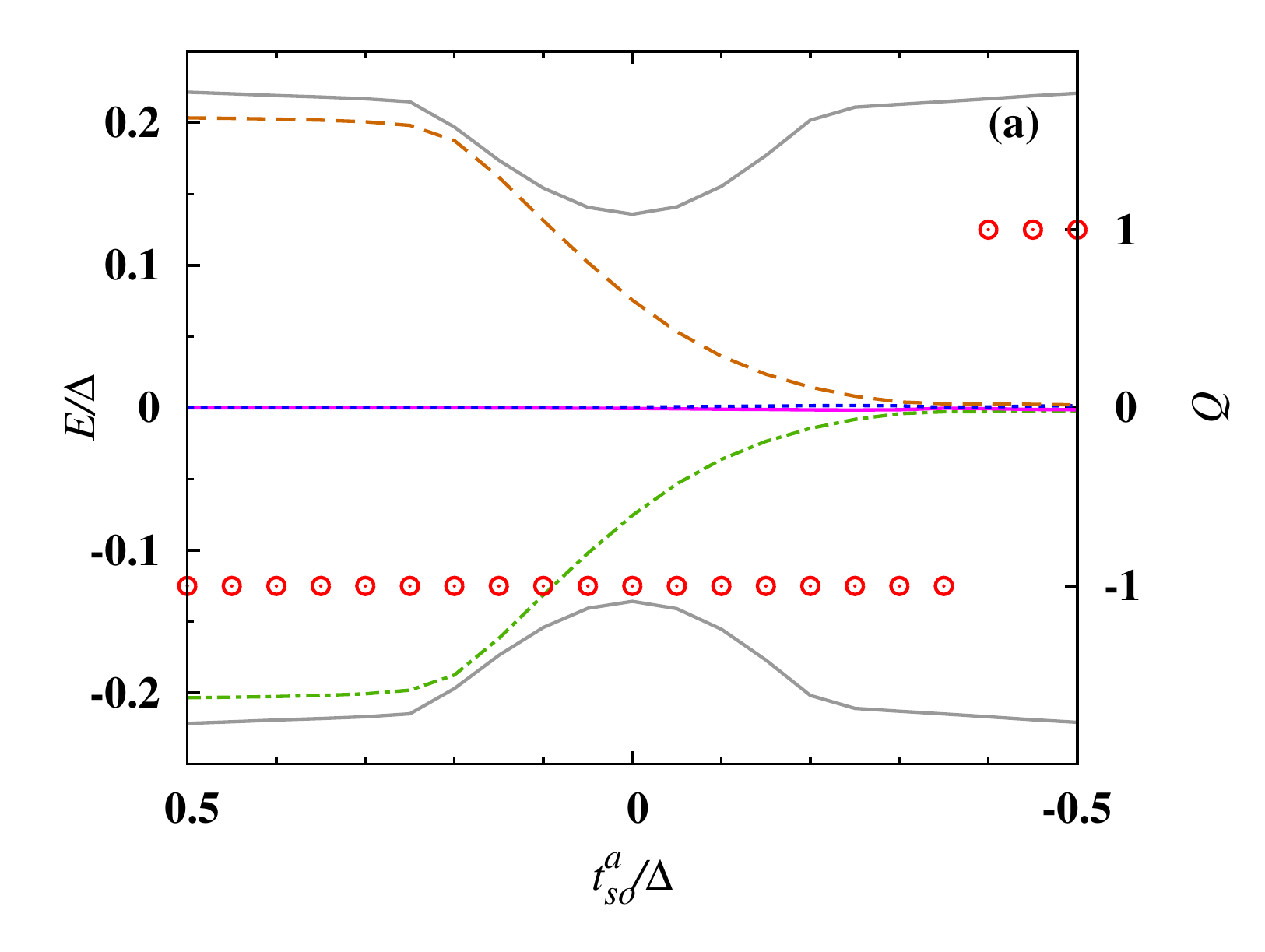}}\\ {\includegraphics[width=9cm]{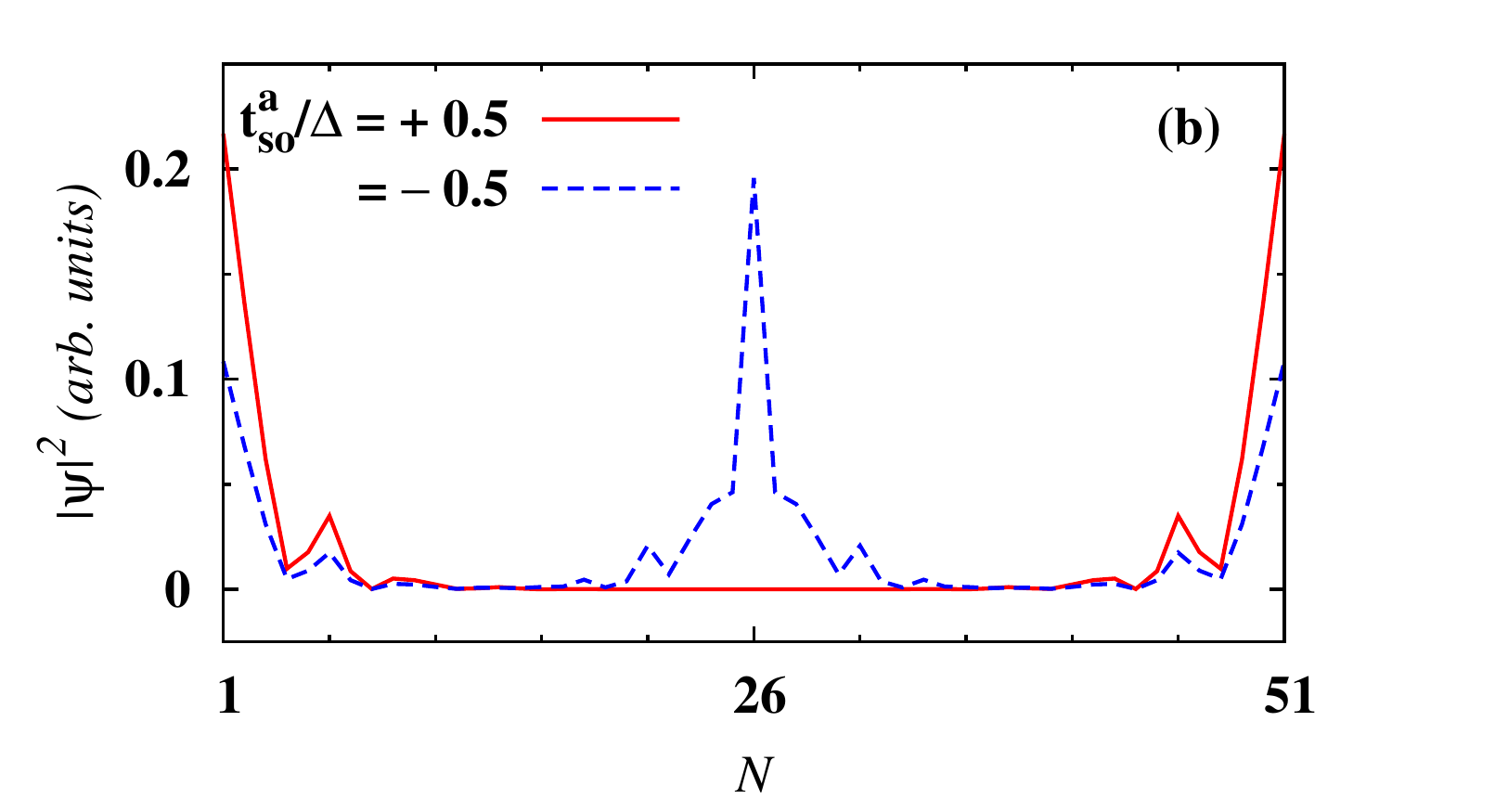}}
  \caption{(Color online) (a) Curves: energy spectrum (with only the lowest six eigenstates close to zero energy plotted)
      in an inhomogeneous QD chain with a finite length (in the calculation
      we set the total number $N$ of QDs to be 51), versus the variation of spin-flip hopping in one half of the QD
      chain $t^a_{\rm so}$. Circles: the topological quantum number $Q$ [in
        Eq.~(\ref{q})] of this inhomogeneous QD chain (with the scale on the
      right-hand side of the frame), versus the variation of spin-flip hopping in one half of the QD
      chain $t^a_{\rm so}$. The spin-flip hopping in the other half of the QD chain remains
      invariant as $t_{\rm so}=0.5\Delta$. (b) Square of the wave function $|\Psi|^2$ of the
      state with its energy closest to zero. The solid curve stands for the
      weakly-coupled MBS in a homogeneous QD chain where $t_{\rm so}=t^a_{\rm
        so}=0.5\Delta$, while the dashed curve stands for the state where the
      MBS have disappeared due to their coupling to the interface
      fermionic bound states in an inhomogeneous QD chain. For the homogeneous
      QD chain, $t_{\rm so}=t^a_{\rm so}=0.5\Delta$; while for the inhomogeneous
      QD chain: $t_{\rm so}=-t^a_{\rm so}=0.5\Delta$.} 
   \label{fig4}
\end{figure}

\subsection{Inhomogeneous QD chain with disordered spin-flip hopping}
We now focus on the robustness of the topological phase against disorder in the 
spin-flip hopping. Again, for simplicity, we assume that due to disorder, the spin-flip hopping
fluctuates in an interval $(t_{\rm so}-\delta_{t_{\rm so}},t_{\rm so}+\delta_{t_{\rm
    so}})$ with a uniform distribution. We find that the topological phase is
relatively robust against disorder in the spin-flip hopping, as long as the spin-flip hopping is
of uniform sign (i.e., $\delta_{t_{\rm so}}<t_{\rm so}$). Nevertheless, when
disorder induces sign-flip in the spin-flip hopping ($\delta_{t_{\rm so}}>t_{\rm so}$), a
topological-nontopological phase transition in the QD chain takes place in the low-chemical-potential region. This feature
can be observed from Fig.~\ref{fig3}, which presents the phase diagrams of 
disordered QD chains with increasing $\delta_{t_{\rm so}}$.

When the spin-flip hopping changes sign along the QD chain, a pair of zero-energy fermionic bound
states\cite{Klinovaja62} arise at the interface between the neighboring domains with different
signs of the spin-flip hopping. These interface fermionic bound states can couple to
other nearby bound states, including the MBS originally present at the ends of the
QD chain. These couplings can destroy the zero-energy MBS. To obtain a clear view of
the interface fermionic bound states and their coupling to the MBS, we further
consider a simple case where a short QD chain possesses a constant spin-flip
hopping on one half of the chain but a varying spin-flip hopping on the other
half. Typically, we study a chain with 51 QDs connected by s-wave
SCs. We set the spin-flip hopping between the neighboring QDs from
the 1st to 26th sites as a constant $t_{\rm so}$, and adjust from $t_{\rm so}$ to $-t_{\rm
  so}$ the spin-flip hopping $t_{\rm so}^a$ on the remaining part. The curves in
Fig.~\ref{fig4}(a) show the energy spectrum of such an
inhomogeneous system (the lowest six eigenstates close to zero are plotted) versus the parameter $t_{\rm so}^a$. It is clearly shown
that with the decrease and eventually the sign-flip of $t_{\rm so}^a$, the
bulk gap in the QD chain gradually closes and the zero-energy fermionic bound
states located around the 26th QD arise. Accordingly, the topological quantum
number $Q$ changes from $-1$ to 1 [as shown by the open circles in
  Fig.~\ref{fig4}(a)], indicating the disappearance of the MBS due to
their coupling to the fermionic bound states. In Fig.~\ref{fig4}(b), we further
present the square of the wave function of the lowest eigenstate, for the cases
with $t_{\rm so}^a=t_{\rm so}$ and $t_{\rm so}^a=-t_{\rm so}$. It is found that
when $t_{\rm so}^a=t_{\rm so}$, i.e., the QD chain is homogeneous, two weakly-coupled MBS are present. However, when $t_{\rm so}^a=-t_{\rm so}$, a state
resulting from the coupling between MBS and the interface bound state replaces the
original MBS.

\section{conclusion}
In this work, we have studied the MBS in a disordered QD chain in proximity to an s-wave
SC. We describe this one-dimensional system by a tight-binding model. By calculating the topological quantum number based on the
scattering-matrix method, we can identify the topological property of such a
QD chain. In our study, we take into account disorder in both the spin-independent terms
(including the chemical potential and the regular spin-conserving hopping) and
the spin-independent term, i.e., the spin-flip hopping due to the Rashba SOC.

We find that the MBS are {\it not} completely immune to disorder in the spin-independent terms, especially when the disorder is
strong. Meanwhile, the MBS are relatively robust against
disorder in the spin-flip hopping, as long as the spin-flip hopping is
of uniform sign. Nevertheless, when the disorder induces sign-flip in spin-flip hopping, a topological-nontopological phase transition in the quantum
dot chain takes place in the low-chemical-potential region. This study
may provide insight into the search of MBS in solid-state systems. 

\begin{acknowledgments}
The authors gratefully acknowledge E. Ya. Sherman and \.{I}. Adagideli for valuable discussions and comments.
P.Z. acknowledges the support of a JSPS Foreign Postdoctoral Fellowship under Grant
No. P14330. F.N. is partially supported by the RIKEN iTHES Project, MURI Center
for Dynamic Magneto-Optics via the AFOSR award number FA9550-14-1-0040, the IMPACT program of JST, and a Grant-in-Aid for Scientific Research (A).
\end{acknowledgments}

\begin{appendix}
\section{Numerical method}
\label{ap1}
As shown in Sec.~\ref{method}, the topological quantum number $Q$ is determined by the
reflection matrix $R$, which can be obtained by the transfer matrix ${M}$ via
Eq.~(\ref{R}). However, the recursive construction [i.e., Eq.~(\ref{tildem})] is numerically
unstable.\cite{Choy195442,snyman045118} We stabilize it by using the method described in
Ref.~\onlinecite{snyman045118}. We briefly introduce this process here.

We denote
\begin{align}
{M}_n=\left(
\begin{array}{cc}
  a_n & b_n\\
  c_n & d_n
\end{array}\right)
\end{align}
and define
\begin{align}
{\cal M}_n=\left(
\begin{array}{cc}
  {\cal A}_n & {\cal B}_n\\
  {\cal C}_n & {\cal D}_n
\end{array}\right)={M}_n{M}_{n-1}...{M}_2{M}_{1}.\label{eq1}
\end{align}
Here $\{a_n, b_n, c_n, d_n\}$ and $\{{\cal A}_n, {\cal B}_n, {\cal C}_n, {\cal D}_n\}$ are $4\times 4$ subblock
matrices. In such framework, ${M}={\cal M}_N$. Further, according to
Eq.~(\ref{R}), we have $R=-{\cal D}_N^{-1}{\cal C}_N$
and $T={\cal A}_N-{\cal B}_N{\cal D}_N^{-1}{\cal C}_N$.

Based on Eqs.~(\ref{mn}) and (\ref{trans}), one finds that
\begin{align}
{M}_n^\dagger\Sigma_z{M}_n=\Sigma_z,\:\Sigma_z=\left(
\begin{array}{cc}
  I & 0\\
 0 &  -I
\end{array}
\right).\label{current}
\end{align}
Therefore, one can construct a unitary matrix $W_n$ from the non-unitary matrix ${M}_n$ as 
\begin{align}
W_n=\left(
\begin{array}{cc}
  u_n & v_n\\
  r_n & s_n
\end{array}\right)=\left(
\begin{array}{cc}
  -d_n^{-1}c_n & d_n^{-1}\\
  a_n-b_nd_n^{-1}c_n & b_nd_n^{-1}
\end{array}\right).\label{eq2}
\end{align}
Now let us define
\begin{align}
{\cal W}_n=\left(
\begin{array}{cc}
 {\cal U}_n & {\cal V}_n\\
  {\cal R}_n & {\cal S}_n
\end{array}\right)=W_n\odot W_{n-1}...W_2\odot W_1\label{eq4},
\end{align}
where the operation $\odot$ is performed as 
\begin{align}\nonumber
&\left(
\begin{array}{cc}
  u_2 & v_2\\
  r_2 & s_2
\end{array}
\right)
\odot
\left(
\begin{array}{cc}
  u_1 & v_1\\
  r_1 & s_1
\end{array}
\right)
\\&=\left(
\begin{array}{cc}
 u_1+v_1(1-u_2s_1)^{-1}u_2r_1 & v_1(1-u_2s_1)^{-1}v_2\\
 r_2(1-s_1u_2)^{-1}r_1 & s_2+r_2(1-s_1u_2)^{-1}s_1v_2 
\end{array}
\right).
\end{align}
In this way, ${\cal W}_n$ is the unitary counterpart of ${\cal  M}_n$, i.e.,
\begin{align}
\left(
\begin{array}{cc}
  {\cal U}_n & {\cal V}_n\\
  {\cal R}_n & {\cal S}_n
\end{array}\right)=\left(
\begin{array}{cc}
  -{\cal D}_n^{-1}{\cal C}_n & {\cal D}_n^{-1}\\
  {\cal A}_n-{\cal B}_n{\cal D}_n^{-1}{\cal C}_n & {\cal B}_n{\cal D}_n^{-1}
\end{array}\right).\label{eq3}
\end{align}
As a result, for numerical stability, instead of calculating ${\cal M}_n$ by
Eq.~(\ref{eq1}), one can calculate the unitary matrix ${\cal W}_n$ based on Eq.~(\ref{eq4}). 

Finally, the topological quantum number $Q$ can be obtained via the relation
\begin{align}
  Q={\rm sgn\:Det}(R)={\rm sgn\:Det}(-{\cal D}_N^{-1}{\cal C}_N)={\rm
    sgn\:Det}({\cal U}_N).
  \end{align}
\end{appendix}

\end{document}